\documentclass[10pt,letter]{article}

\usepackage{hyperref}
\usepackage[authoryear]{natbib}
\usepackage{times}

\usepackage{mathrsfs}
\usepackage{graphicx}
\def\cssm{\textsl{CytoSaddleSum}}
\def\ssum{\textsl{SaddleSum}}

\begin{document}

\begin{titlepage}

\begin{center}
{\Large\bf {CytoSaddleSum}: a functional enrichment analysis plugin for {Cytoscape} based on sum-of-weights scores}
\end{center}
\vspace{.35cm}

\begin{center}
{\large Aleksandar Stojmirovi\'c\,, Alexander Bliskovsky and Yi-Kuo Yu\footnote{to whom correspondence should be addressed}}
\vspace{0.25cm}
\small

\par \vskip .2in \noindent
National Center for Biotechnology Information\\
National Library of Medicine\\
National Institutes of Health\\
Bethesda, MD 20894\\
United States
\end{center}

\normalsize
\vspace{0.25cm}

\begin{abstract}

\subsubsection*{Summary:}
\cssm\ provides Cytoscape users with access to the functionality of \ssum,
a functional enrichment tool based on sum-of-weight scores. 
It operates by querying \ssum\ locally (using the
standalone version) or remotely (through an HTTP request to a web server). The
 functional enrichment results are shown as a term relationship
network, where nodes represent terms and edges show term
relationships. Furthermore, query results are written as Cytoscape attributes
allowing easy saving, retrieval and integration into network-based data analysis workflows.

\subsubsection*{Availability:}
\href{http://www.ncbi.nlm.nih.gov/CBBresearch/qmbp/Yu/downloads}{www.ncbi.nlm.nih.gov/CBBresearch/Yu/downloads} \\
The source code is placed in Public Domain.

\subsubsection*{Contact:} \href{yyu@ncbi.nlm.nih.gov}{yyu@ncbi.nlm.nih.gov}
\end{abstract}
\end{titlepage}

\section{Introduction}

\cssm\ is a Cytoscape~\citep{SORW11} plugin to access the functionality of \ssum, an enrichment analysis tool based on sum-of-weights-score~\citep{SY10a}. Unlike most other enrichment tools, \ssum\ does not require users to directly select significant genes or perform extensive simulations to compute statistics. Instead, it uses weights derived from measurements, such as log-expression ratios, to produce a score for each database term. It then estimates, depending on the number of genes involved, the P-value for that score by using the saddlepoint approximation~\citep{LR80} to the empirical distribution function derived from all weights. 
This approach was shown~\citep{SY10a} to yield accurate P-values and internally consistent retrievals.  

As a popular and flexible platform for visualization, integration and analysis of network data, Cytoscape allows gene expression data import and hosts numerous plugins for functional enrichment analysis. However, none of these plugins are based on the `gene set analysis approach' that takes into account gene weights. Therefore, to fill this gap, we have developed \cssm, a Cytoscape interface to \ssum. 
To enable several desirable features of \cssm, however, we had to significantly extend the original 
\ssum\ code (see descriptions below). 

\section{Implementation}\label{sec:implement}

While \cssm\ is implemented in Java using Cytoscape API, it functions by running either locally or remotely a separate  instance of \ssum, written in C. In either mode, \cssm\ takes the user input through a graphical user interface, validates it, and passes a query to \ssum. Upon receiving the entire query results, 
\cssm\ stores them as the node and network attributes of the newly-created term relationship graph.  
 Consequently, the query output can be edited or manipulated within Cytoscape. 
Furthermore, saving term graph through Cytoscape also preserves the results for later use. 

 The most important extension to \ssum\ involved construction of extended term databases (ETDs). Each ETD contains the mappings of genes to Gene Ontology~\citep{GeneOnt10} terms and KEGG~\citep{KAGH08} pathways, as well as an abbreviated version of the NCBI Gene~\citep{MOPT11} database for all genes mapped to terms. Thanks to the latter, when using an ETD, \ssum\ is able to interpret the provided gene labels as NCBI Gene IDs, as gene symbols and as gene aliases.
Each ETD also contains relations among terms that are used by \ssum\ for term graph construction.

\begin{figure*}[t!]
\begin{center}
\scalebox{0.5}{\includegraphics{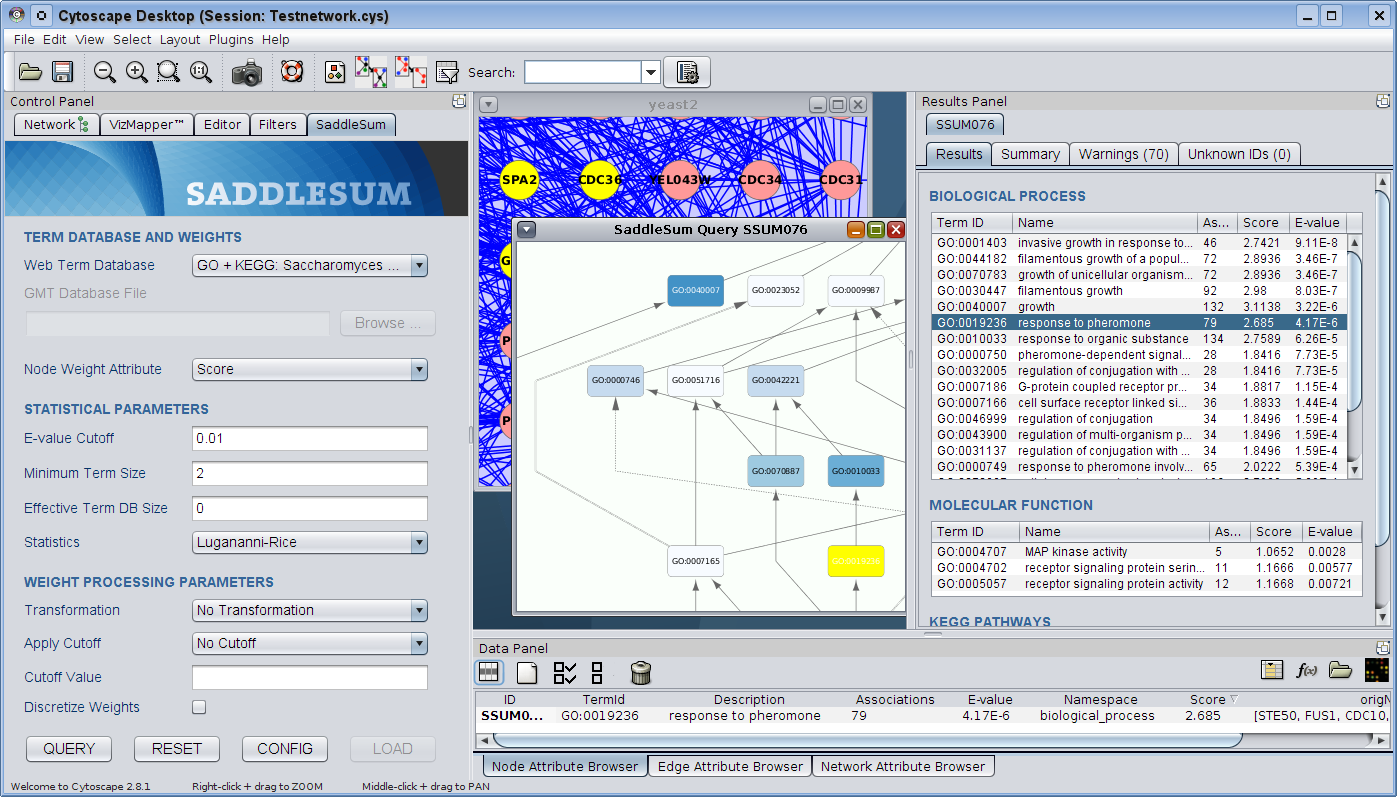}}
\caption{\cssm\ user interface consists of the query form (left), the results panel (right) and the term relationship network (center), which here partially covers the original network. The results stored as attributes of the term network can be edited through Cytoscape Data Panel.} \label{fig:screenshot}
\end{center}
\end{figure*}

\section{Usage}\label{sec:usage}
\cssm\ operates on the currently selected Cytoscape network whose nodes represent genes or gene products. The queries are submitted through the query form embedded as a tab into the Cytoscape Control Panel, on the left of the screen. The selected network must contain at least one node mapped to a floating-point Cytoscape attribute, which would provide node weights. \cssm\ considers only the selected nodes within the network.
 The user can select the weight attribute through a dropdown box on the query form.
Any selected node without specified weight is assumed to have weight 0. 
 The user-settable \textit{cannonicalName} attribute, automatically created by Cytoscape for each network node, serves as the gene label.  

After selecting the network and the nodes within it, the user needs to select a term database and set the statistical and weight processing parameters. The latter enable users to transform the supplied weights within \ssum. 
This includes changing the sign of the weights, as well as applying a cutoff, by weight or by rank. All weights below the cutoff are set to 0. The statistical parameters are E-value cutoff, minimum term size, effective database size and statistical method. 
We define the effective database size as the number of terms in the term database that map to at least $k$ genes among the selected nodes, where $k$ is the minimum term size. 
Apart from the default `Lugannani-Rice' statistics, it is also possible to select `One-sided Fisher's Exact test' statistics, which are based on the hypergeometric distribution. In that case, the user must select a cutoff under the weight processing parameters.

To run local queries, a user needs the command-line version of \ssum\ and the term databases, both available for download from our website, and install them on the same machine that runs Cytoscape. The advantages of running local queries include speed, independence of Internet connection, and support of queries to custom databases in the GMT file format used by the GSEA tool~\citep{STMM05}.
Furthermore, the standalone program can be used outside of Cytoscape for large sets of queries. On the other hand, running remote queries require no installation of additional software, since queries are passed to the \ssum\ server over an HTTP connection.  The disadvantage of running remote queries is that it can take much longer to run and that the choice of term databases is restricted to ETDs available only for some model organisms.

\cssm\  also displays warning or error messages reported by \ssum. For example, 
when a provided gene label is ambiguous, depending on whether the ambiguity could be resolved, \cssm\ will relay 
a warning or an error message reported by \ssum. 
\cssm\ presents query results as a term relationship network (Fig.~\ref{fig:screenshot}), consisting of significant terms or their ancestors linked by hierarchical relations available in the term database. The statistical significance of each term is indicated by the color of its corresponding node. To facilitate browsing of the results, \cssm\ generates a set of summary tables, which contain the lists of significant terms and various details about the query. These summary tables are embedded into Cytoscape Results Panel, on the right of the screen. Clicking on a significant term in a summary table will select that term in the term relationship network and select all nodes mapping to it in the original network. The results can be exported as text or tab-delimited files and can be restored from tab-delimited files through the Export and Import menus of Cytoscape.Detailed instructions and explanations can be found in \ssum\ manual available from our website.

\section*{Acknowledgments}
This work was supported by the Intramural Research Program of the National Library of Medicine at National Institutes of Health.


\end{document}